\documentclass{article}
\usepackage{spconf,amsmath,graphicx}
\usepackage[font=small,skip=0pt]{caption}
\usepackage{amsfonts}
\usepackage{wrapfig}
\usepackage{booktabs}

\title{Noise Robust TTS for Low Resource Speakers using Pre-trained Model and Speech Enhancement}

\name{\begin{tabular}{c}Dongyang Dai$^{1,2,\dagger,\#}$  \thanks{$\dagger$ Equal contribution}\thanks{\# This work is done during internship at ByteDance} \thanks{Audio samples: https://noisetaco.github.io/noisetaco}, Li Chen$^{1,\dagger}$, Yuping Wang$^1$ , Mu Wang$^1$, \\ Rui Xia$^1$, Xuchen Song$^1$, Zhiyong Wu$^2$, Yuxuan Wang$^1$\end{tabular}}

 \address{ $^1$ByteDance, Inc. \\
 	$^2$Tsinghua-CUHK Joint Research Center for Media Sciences, Technologies and Systems, \\
 	Graduate School at Shenzhen, Tsinghua University, Shenzhen, China\\
 	\small{\{daidongyang,chenli.cloud,wangyuping,rui.xia,xuchen.song,wangyuxuan.11\}@bytedance.com, zywu@se.cuhk.edu.hk}}
 
\begin{document}

\ninept
\maketitle
\begin{abstract}
With the popularity of deep neural network, speech synthesis task has achieved significant improvements based on the end-to-end encoder decoder framework in the recent days. More and more applications relying on speech synthesis technology have been widely used in our daily life. Robust speech synthesis model depends on high quality and customized data which needs lots of collecting efforts. It is worth investigating how to take advantage of low-quality and low resource voice data which can be easily obtained from the Internet to synthesize personalized voice. In this paper, the proposed end-to-end speech synthesis model uses both speaker embedding and noise representation as conditional inputs to model speaker and noise information respectively. Firstly, the speech synthesis model is pre-trained with both multi-speaker clean data and noisy augmented data; then the pre-trained model is adapted on noisy low-resource new speaker data; finally, by setting the clean speech condition, the model can synthesize the new speaker’s clean voice. Experimental results show that the speech generated by the proposed approach has better subjective evaluation results than the method directly fine-tuning pre-trained multi-speaker speech synthesis model with denoised new speaker data. 
\end{abstract}
\begin{keywords}
Noise Robust TTS, Mel-spectrogram Denoise Masks, Low Resourse Personalized TTS
\end{keywords}
\section{Introduction}

Text-to-speech (TTS) technology has been widely used in many products, such as e-books, voice assistants, automatic navigation, etc. Recently, with the development of neural networks, end-to-end TTS models, such as Tacotron \cite{wang2017tacotron},   Char2Wav \cite{sotelo2017char2wav}, DeepVoice3 \cite{ping2017deep} and Tacotron2 \cite{shen2018natural} have gradually become mainstream. End-to-end TTS models using an attention based encoder-decoder structure, learning patterns from large amount of data, can produce more natural sound than traditional parametric TTS systems \cite{zen2009statistical}.

Based on the end-to-end model, many researchers have begun to pay attention to how to control the style, tone, and other information of synthesized speech.  \cite{wang2018style} proposed Global Style Tokens to represent the speech's style information.  \cite{zhang2019learning} applied variational autoencoder (VAE) to model the distribution of speech's style features. As for personalized TTS,  \cite{NIPS20188206} introduced speaker adaptation and speaker encoding approaches for voice cloning.  As demonstrated in \cite{NIPS20188206}, speaker adaptation approach, which is based on fine-tuning a pre-trained multi-speaker model for an unseen speaker using a few samples, achieves better performance. Given the abundance of audio and video information proliferating on the Internet, finding effective ways to synthesize wide variety of personalized sound using widely available low-quality voice data has become an interesting topic.

Noise Robust TTS \textendash- training a stable TTS model on noisy, low-quality data, has long been of interest to researchers in the field.  Authors in \cite{valentini2016investigating} introduced speech enhancement methods for noise robust TTS.  In their solution, an recursive neural network (RNN) based speech enhancement model is applied to map acoustic features extracted from noisy speech to features describing clean speech; the enhanced data is then used to adapt a pre-trained hidden Markov model (HMM) based TTS acoustic model; finally, STRAIGHT \cite{kawahara1999restructuring} vocoder is used to generate waveform from acoustic features. However, the speaker information will more or less been reduced by preprocessing through speech enhancement model.  Besides, due to better effects in terms of sound quality and naturalness, neural network-based acoustic models and vocoders have replaced HMM models and STRAIGHT as mainstream.

To leverage low-quality crowd-sourced data to train multi-speaker TTS models that can synthesize clean speech for all speakers,  based on Tacotron2, \cite{hsu2019disentangling} introduced conditional generative reference encoders and adversarial training to learn disentangled representations to independently control the speaker identity and background noise in generated signals.  The method applies speaker encoder to learn speaker related variable $z_s$, and residual encoder to extract variable $z_r$ to model unlabelled attributes (e.g. acoustic conditions). As the speaker encoder is followed by speaker classifier and gradient reversal layer followed with augmentation classifier, $z_s$ respresents noise-free speaker related information; $z_r$ represents noise related infomration. However, the noise signal is not always stable, which varies along with time. Therefore a fixed-length vector is not enough to model the noise information, especially in the case of speech data with low signal-to-noise ratio (SNR).

Recently, neural network-based methods using masks have achieved excellent results in tasks such as speech enhancement and speech separation, and have become mainstream \cite{narayanan2013ideal,wang2014training}. Inspired by these works, we use variable-length Mel-spectrogram denoise masks instead of a fixed-length vector as the representation of noise information. We assumed that the noisy speech is generated by the noise signal adding to clean speech, so for each point in the Mel frequency domain, the energy $ E $ contains the energy of the clean speech $ E_s $, and the energy of the noise signal $ E_n $. So the value of the Mel-spectrogram denoise mask of the corresponding point is calculated by $E_s / E$. In this paper, based on voice cloning framework, we adopt the Mel-spectrogram denoise masks (including noisy masks for noisy speechs and clean masks for clean speechs) as noise representation which conditions on the end-to-end speech synthesis model, and pre-train the TTS model on multi-speaker's enhancement data for noise robust personalized TTS. Our work is summarized in the following three aspects:

\begin{enumerate}
	\item The proposed method uses both the speaker embedding and the noise representation as conditional inputs of the basic end-to-end speech synthesis model to achieve independent control of the synthesized speech with noise and different speakers.
	After being pre-trained on multi-speaker enhancement data, model is adapted on low-quality new speaker's data and can synthesize clean speech for the new speaker.
	\item The proposed model uses Mel-spectrogram denoise masks as the noise representation. Compared with a fixed-length vector, Mel-spectrogram denoise mask can better characterize the noise information. The model is pre-trained on both noisy and clean multi-speaker data. The noise representation includes noisy masks (extracted from noisy speech) and clean masks (all elements equal to 1). The model accepts noisy masks as conditional input to generate speech with corresponding noise, and similarly accepts clean masks to generate clean speech. When the noise representation varies from noisy masks to clean masks, the generated speech changes from noisy to clean.
	\item The proposed model accepts the features extracted by the pre-trained speaker recognition model as the speaker embedding, and the TTS model is pre-trained on multi-speaker voice data, which realizes personalized speech synthesis for low-quality low-resource new unseen speaker's data.
\end{enumerate}

\section{Methodology}

\subsection{Proposed Approach}

\begin{figure}[t]
	\centering
	\includegraphics[scale=0.32]{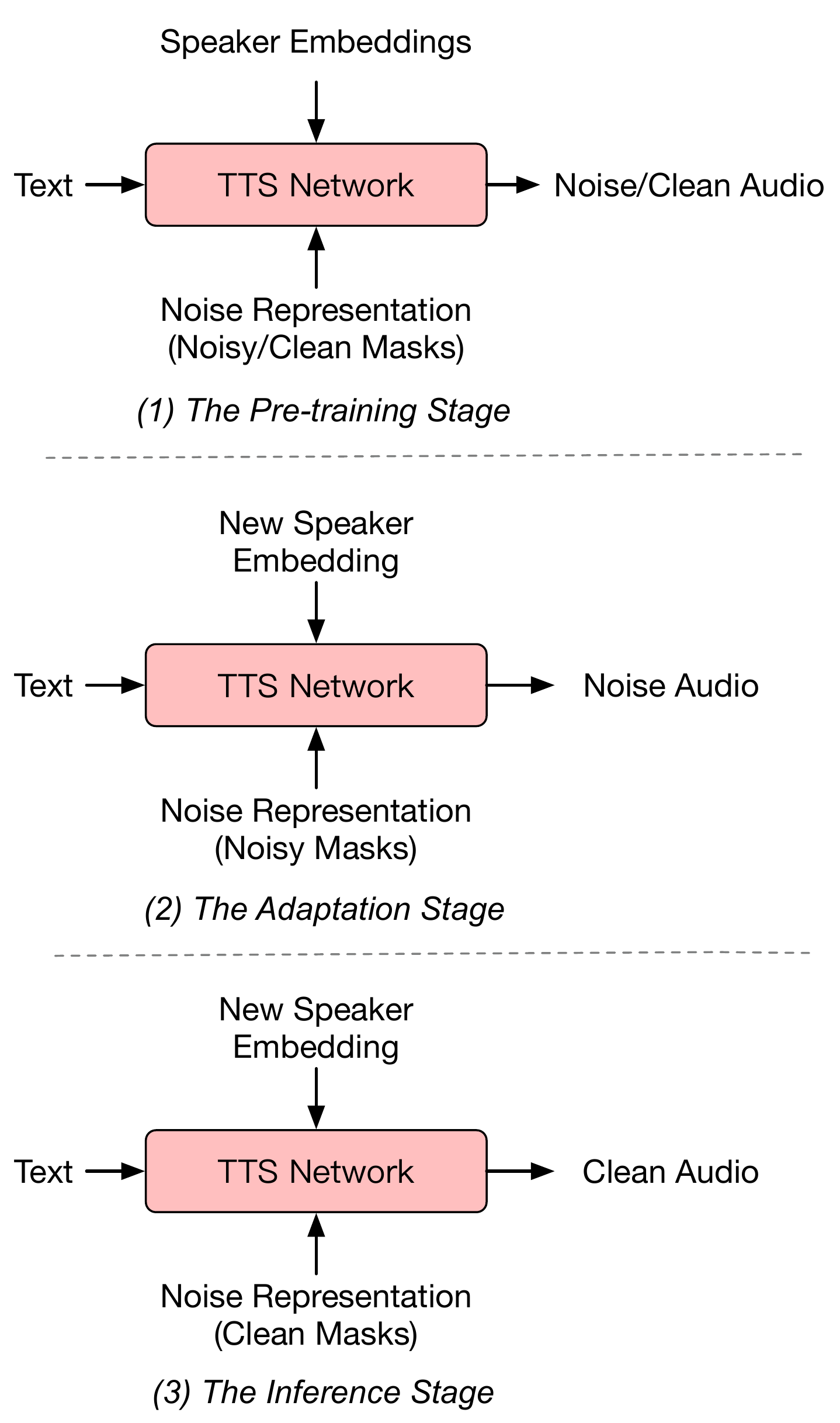}
	\caption{The proposed approach}
	\label{fig:approach}
	\vspace{-10pt}
\end{figure}

As depicted in Figure \ref{fig:approach}, the proposed approach consists of three stages:
\begin{enumerate}
	\item The pre-training stage: The end-to-end TTS model is pre-trained on clean and noisy multi-speaker voice data. The model accepts speaker embedding and noise representation as conditional inputs. The speaker embedding is extracted via a speaker recognition model, and the noise representation is the Mel-spectrogram denoise masks, including noisy masks and clean masks. The noisy masks is the predicted Mel-spectrogram denoise masks extracted by a speech enhancement model from noisy speech, while the clean masks is corresponding to clean speech whose value all equals to 1.
	\item The adaptation stage: The pre-trained model is adapted on the new low-quality low-resource speaker data. The new speaker data only contains noisy speech, so the noise representation only contains noisy masks.
	\item The inference stage: The adapted model accepts clean masks as conditional input to synthesize clean voice of the new speaker. Where clean masks represents the Mel-spectrogram denoise masks of  clean speech, and the value of each element is set to 1.
\end{enumerate}

The following sections will describe the speaker embedding and noise representation extraction, as well as the details of the model. 

\subsection{Speaker Embedding Extraction}
\label{sec:spkembextract}

In order to preserve the relative relationship between different speakers and deal with unseen speakers more conveniently, instead of using one-hot encoding directly, the speaker embedding is extracted by a speaker recognition model, which is pre-trainded on an internal dataset containing about 20,000 speakers. We apply the approach in \cite{xie2019utterance} to implement the speaker recognition model, which consists of a modified ResNet \cite{he2016deep} in a fully convolutional way to extract frame-level features and a following GhostVLAD \cite{zhong2018ghostvlad} layer for feature aggregation along the temporal axis. To extract more discriminative speaker embedding, additive margin softmax \cite{wang2018additive} is adopted to train the speaker recognition model. Besides, all the speaker embeddings is normalized to the unit hypersphere space. In practice, we adopt the center point of utterance-level speaker embeddings corresponding to the same speaker as the speaker-level speaker embedding, and the speaker-level speaker embedding will be fed into TTS model. On unit hypersphere space, the center is calculated by $\boldsymbol\mu = \sum \boldsymbol x_i / || \sum \boldsymbol x_i ||$, where $\boldsymbol\mu$ is the speaker-level speaker embedding and $\boldsymbol x_i$ is the $i$-th utterance-level speaker embedding.

\subsection{Noise Representation Extraction}
\label{ch:spkembextrac}

Inspired by recent advancement in speech enhancemet \cite{narayanan2013ideal,valin2018hybrid}, we apply Mel-spectrogram denoise masks as noise representation to model the noise information. The Mel-spectrogram denoise masks is extracted by a speech enhancement model which accepts Mel-spectrogram as input, the model structure is a variant of the CNN-RNN-FC structure, and it is very similar to the model in \cite{valin2018hybrid}.  Compared with \cite{valin2018hybrid}, the proposed model has following three main differences:
\begin{enumerate}
	\item The input feature of the model is Mel-spectrogram instead of linear spectrogram.
	\item The LSTM layer is replaced with DFSMN(Deep Feedforward Sequential Memory Networks) \cite{zhang2018deep}, which can reduce the model size (from about 26M parameters to about 4.76M). Meanwhile, the inference speed is improved by about 9x due to the reason that computation of DFSMN layer can be parallelized (speed testing is benchmarked on Nvidia T1080).
	\item The activation function of the output layer is sigmoid, and its output is assumed to be the mask $M$. Since mean square error (MSE) is enough to extract effective Mel-spectrogram denoise masks  according to our experiments, we directly use MSE as the model's loss function as shown in Equation \ref{eq:seloss}, where $S_{noise}$ denotes the Mel-spectrogram of noisy speech, $S_{clean}$ denotes the corresponding clean speech's Mel-spectrogram. $\odot$ means element-wise multiply,  $S_{noise} \odot M$ represents the denoised Mel-spectrogram. $n$ is the number of Mel-spectrogram's TF-bins.
\end{enumerate}
\begin{equation}
\label{eq:seloss}
L_{mse} = \frac{\sum{\left\| S_{noise} \odot M - S_{clean}\right\|^2}}{n}
\end{equation}


\subsection{Noise Robust Personalized TTS Model}

\subsubsection{Basic TTS model}

We use an encoder-decoder-based Tacoton-like end-to-end neural network as the basic TTS model. Compared with the original Tacotron \cite{wang2017tacotron}, two improvements have been made. Firstly,  GMM-based attention \cite{battenberg2020location} is applied for improving the stability of the synthesis and reducing bad cases. Secondly, in order to improve the quality of the synthesized sound, we use a neural network-based vocoder to generate sound from Mel-spectrogram just like what Tacotron2 \cite{shen2018natural} does, so the outputs of PostNet is  log-level Mel-spectrogram instead of linear spectrogram, and the model is optimized by both before loss and after loss. More details about the decoder will be described in section \ref{ch:noiserobustmodel}.

\subsubsection{Speaker embedding condition}

\begin{figure}[t]
	\centering
	\includegraphics[scale=0.30]{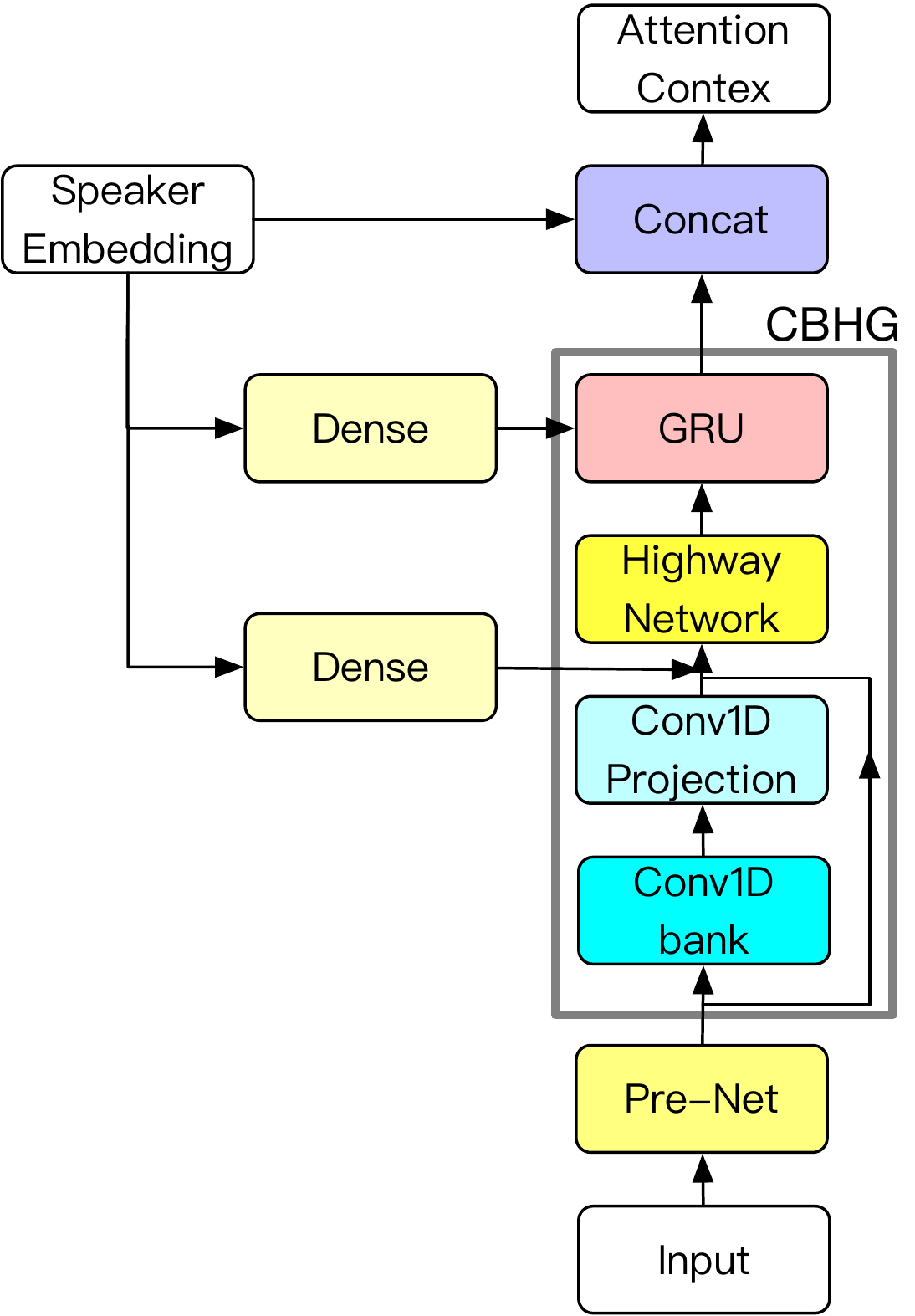}
	\caption{Speaker embedding as the conditional input to encoder}
	\label{fig:SpkEncoder}
	\vspace{-10pt}
\end{figure}

In order to control the speaker characteristics of the synthesized speech, the speaker embedding is taken as the conditional input of the encoder of the TTS model. As shown in Figure \ref{fig:SpkEncoder}, in addition to concatenating with the encoder outputs, the speaker embedding will also pass through two dense layers respectively, then adopted as the additional conditional inputs for Highway Network and the projected speaker embedding will be adopted as the initial value of the GRU layer.

\subsubsection{Noise representation condition}
\label{ch:noiserobustmodel}
\begin{figure}[t]
	\centering
	\includegraphics[scale=0.30]{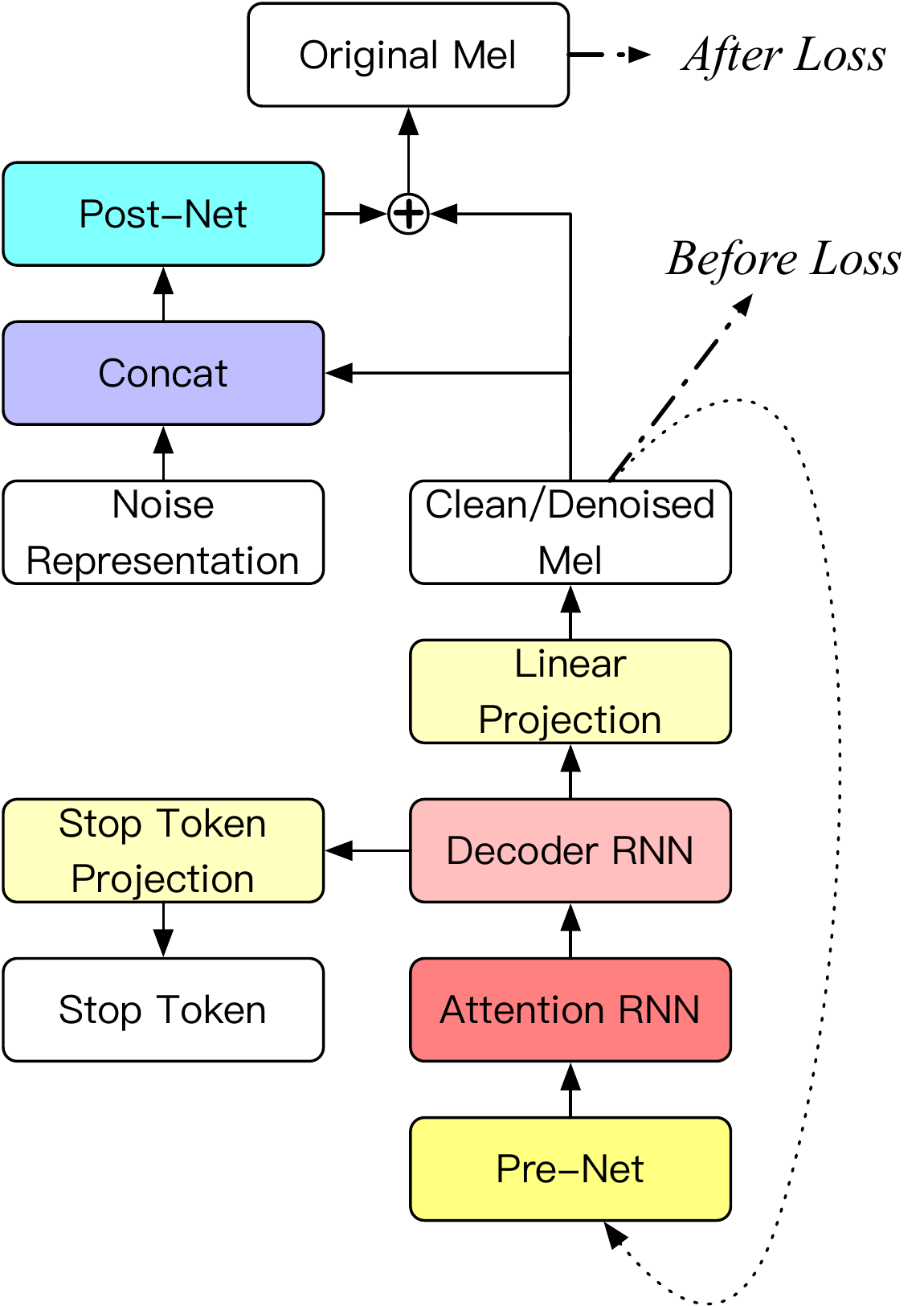}
	\caption{Noise representation as the conditional input to decoder}
	\label{fig:noisedecoder}
	\vspace{-10pt}
\end{figure}
To model the noise information, we use noise representation (Mel-spectrogram denoise masks, extracted from the speech enhancement model) as the conditional input of the decoder. Shown in Figure \ref{fig:noisedecoder}, the decoder is an autoregressive structure, which is composed of Pre-Net, Attention RNN layer, Decoder RNN layer, Linear Projection layer, Stop Token Projection layer and Post-Net. 

In our model,  Post-Net is used to model noise information. In order not to introduce noise information in other parts of the decoder, we use denoised Mel-spectrogram or clean Mel-spectrogram as the target output of the Linear Projection layer to calculate the Before Loss (showed in Figure \ref{fig:noisedecoder}).  The Post-Net accepts both processed noise representation and the output of Linear Projection layer as input, whose outpus is added to the output of Linear Projection layer to predict the original Mel-spectrogram and calculate the After Loss. Since the TTS model outputs log-level Mel-spectrograms, before concatenating to the Linear Projection layer's output as Post-Net's input,  the noise representation is first clipped to a value between 0.1 and 1, then converted to log level, and finally normalized to a value between -4 and 4.

The model is pre-trained on multi-speaker clean data and noisy augmented data, where the noisy augmented speech data is generated from the clean speech data mixed with the noise signal, and each augmented utterance has its corresponding clean speech. So in the pre-training stage, the model calculates the Before Loss according to the clean Mel-spectrogram that is obtained from the corresponding clean speech. And in the adaptation stage, the Before Loss is calculated according to the denoised Mel-spectrogram that is obtained from the speech enhancement model and the corresponding noisy speech.

\section{Experiments and Analysis}

\subsection{Experimental Setup}

First of all, we add the noise signal from Audio Set \cite{45857} to an internal clean multi-speaker TTS corpus to generate a noisy augmented multi-speaker corpus. The proposed model uses both the clean and the noisy augmented corpus during pre-training. The clean multi-speaker TTS corpus contains about 1100 speakers, each speaker has about 30 minutes of voice data. Then, the model is adapted on the low-resouce low-quality new speaker data. In our experiments, four new speaker's data is adopted, each speaker contains 200-300 utterances, and the low-quality data is constructed based on the Microsoft Scalable Noisy Speech Dataset \cite{reddy2019scalable}, ensuring that the SNR of each utterance is less than 5dB.

In order to verify the effect of our proposed model, we use the approach of denoise and then synthesis as baseline. The model of baseline method is the same of the proposed, except that Post-Net only accepts the output of the Linear Projection layer as the input instead of concatenating the Mel-spectrogram denoise masks. The baseline model is only pre-trained on clean multi-speaker data, and the denoised new speaker's Mel-spectrogram processed by the speech enhancement model is used for adaptation. In baseline, we utilized the denoised Mel-spectrogram instead of the original noisy Mel-spectrogram for adaptation as the speech synthesized by the following approach is obviously worse.

At first, a speech enchancement model need to be pretrained for both baseline and the proposed approach. In our experiments, we train the Mel-specdtrogram level enhancement model based on noisy multi-speaker TTS data and its corresponding clean data. Then the Mel-spectrogram denoise masks predicted by the speech enhancement model are used as the condition input of the proposed model, and the denoised Mel-spectrogram obtained by the enhanced model is used for the adaptation of baseline method.

\subsection{Experimental Results and Analysis}

We first verified the effect of the speech enhancement model, which is the basis for noise robust TTS. The speech enhancement model was tested under different SNRs, including -5dB, 0dB and 5dB. Under each SNR setting, there are 61 noisy voices generated with another internal noise sets. The  scale-invariant signal-to-distortion ratio (SI-SDR) \cite{le2019sdr} calculated on Mel-spectrogram is adopted as metric and shown in Table \ref{tab:sesdr}. It can be seen from Table \ref{tab:sesdr} that the speech enhancement model can significantly reduce the noise in speech, especially in the case of low SNR. 

\begin{table}[th]
	\caption{SI-SDR(dB) result of our speech enhancement model on different SNR levels}
	\label{tab:sesdr}
	\centering
	\begin{tabular}{ll}
		\toprule
		\textbf{SNR} & \textbf{SI-SDR}  \\
		\midrule
		$-5$                       & 3.787             \\
		$0$                       & 7.154                \\
		$5$                       & 8.694       \\
		\bottomrule
	\end{tabular}
	\vspace{-10pt}
\end{table}

\begin{figure}[t]
	\centering
	\includegraphics[scale=0.3]{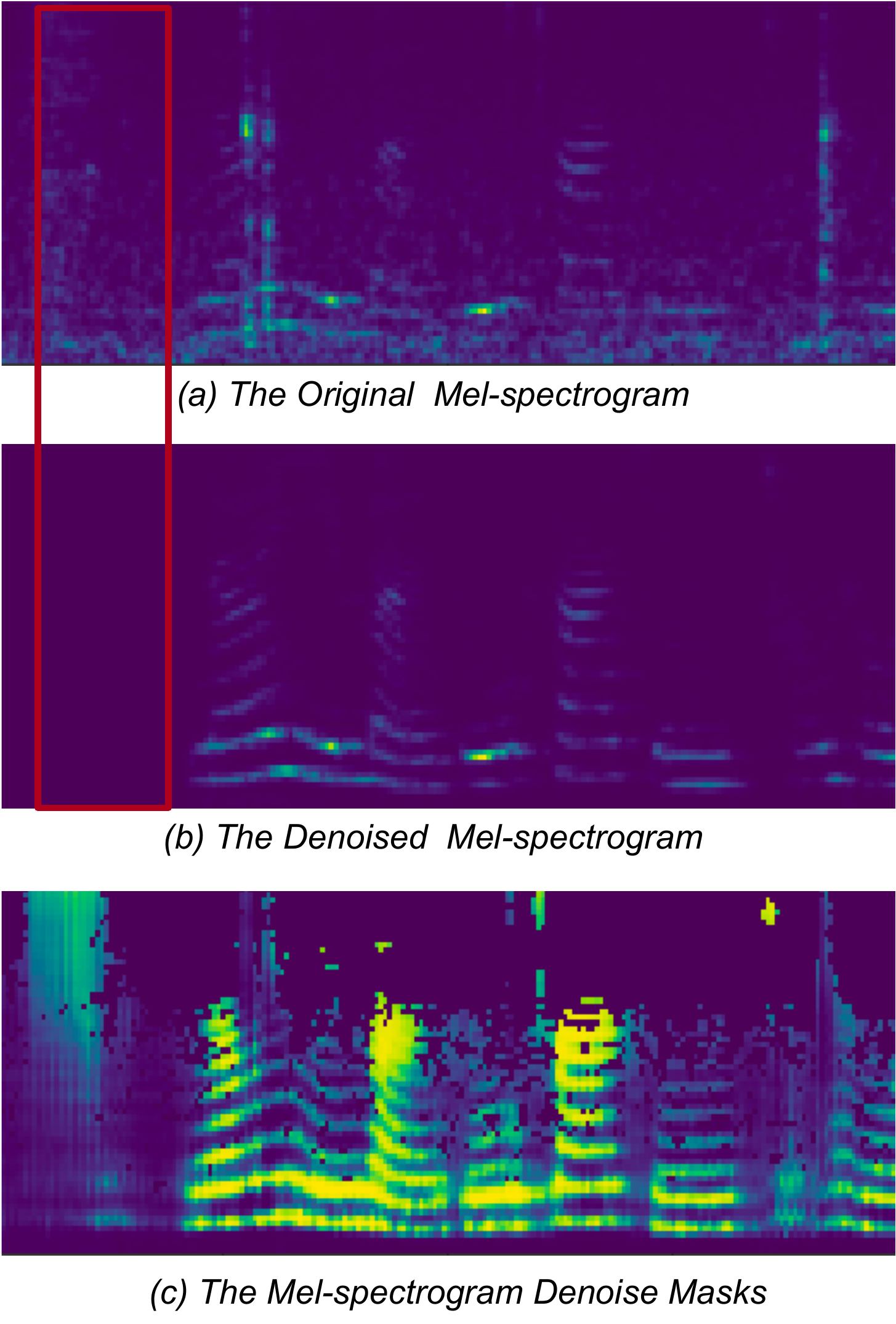}
	\caption{The speech enhancement result and Mel-spectrogram denoise masks}
	\label{fig:enhancementResult}
	\vspace{-10pt}
\end{figure}

An illustration of the enhanced Mel-spectrogram on -5dB utterance is shown in Figure \ref{fig:enhancementResult}. From Figure \ref{fig:enhancementResult}, we can see that some voice information contained in the original Mel-spectrogram was reduced by the speech enhancement model (marked with a red rectangle), indicating that  adapting TTS model directly on the denoised data will result in unstable voice. Comparing Figure \ref{fig:enhancementResult}-(c) with Figure \ref{fig:enhancementResult}-(a) and Figure \ref{fig:enhancementResult}-(b), it can be seen that Mel-spectrogram denoise masks could well represent noise information, which is the reason why we adopt Mel-spectrogram denoise mask (noise representation) as the decoder's condational input.


We synthesized 40 utterances with the same content for 4 new speakers under baseline and proposed approach respectively, meaning 40 x 4 x 2 utterances were synthesized. We conducted mean opinion score tests, where 40 people were asked to evaluate synthesized speech in therm of speech quality. Each person was asked to score randomly selected 40 untterances  from the total 320 ones. The score ranges from 1 to 5, where 1 represents the worst and 5 represents the best.
\begin{table}[th]
	\vspace{-5pt}
	\caption{The MoS result in term of speech quality}
	\label{tab:mos}
	\centering
	\begin{tabular}{lll}
		\toprule
		&\textbf{Baseline} & \textbf{Proposed}  \\
		\midrule
		\textbf{Speaker1}       &       3.018        &    \textbf{3.227}    \\
		\textbf{Speaker2}      &      3.462           &     \textbf{3.560}        \\
		\textbf{Speaker3}       &        3.139       &    \textbf{3.181}    \\
		\textbf{Speaker4}        &       3.313       &        \textbf{3.424}        \\
		\bottomrule
	\end{tabular}
	\vspace{-5pt}
\end{table}

The mean opinion score of each setting is shown in the Table \ref{tab:mos}. It can be seen that with the proposed method, the speech quality of each speaker is improved compared with the baseline. The score value has improved by an average of 0.115. The results demonstrate that our model can synthesize higher quality speech with low-resource low-quality data. We speculate that this is due to the denoised Mel-spectrogram used directly by the baseline loses some voice information, which leads to the instability of speech synthesis.

\begin{table}[th]
	\vspace{-5pt}
	\caption{The cosine similarity of speaker-level speaker embedding from training data and synthesized data}
	\label{tab:spksim}
	\centering
	\begin{tabular}{lll}
		\toprule
		&\textbf{Baseline} & \textbf{Proposed}  \\
		\midrule
		\textbf{Speaker1}       &       0.767        &    0.780    \\
		\textbf{Speaker2}      &      0.826           &     0.829       \\
		\textbf{Speaker3}       &        0.905      &    0.870    \\
		\textbf{Speaker4}        &       0.890      &        0.891      \\
		\bottomrule
	\end{tabular}
	\vspace{-5pt}
\end{table}

For personalized speech synthesis, the timbre similarity between the synthesized speech and the target speech is also an important indicator to measure the system. Therefore, as in section~\ref{ch:spkembextrac}, we extract speaker-level speaker embedding from speech generated by baseline and the proposed respectively, and then calculate the cosine similarity with the corresponding speaker-level speaker embedding from training data to measure the timbre similarity between the synthesized speech and the training data. The results are shown in the Table~\ref{tab:spksim}. In our speech speaker recognition model, if the cosine similarity is greater than 0.70, it is considered the same speaker. Therefore, it shows that the method based on speech enhancement (both baseline and the proposed) can generate speech whose timbre close to the target.


\section{Conclusion}

In this paper, we propose a novel method for synthesizing personalized speech based on the end-to-end network model in the case of low quality and low resources data. The model accepts speaker embedding and Mel-spectrogram denoise mask as the conditional input, for modeling speaker and noise information respectively. The model is first pre-trained on clean multi-speaker data and augmented noisy multi-speaker data,  then adapted on the low-resource low-quality new-speaker data, and finally utilized to synthesize clean voices of the new speaker. Experimental and subjective evaluation results show that the proposed approach can synthesize better speech compared to baseline method, which fine-tunes the pre-trained multi-speaker TTS model on the denoised new speaker’s data directly.

\bibliographystyle{IEEEbib}
\bibliography{mybib}

\begin{thebibliography}{1}

\bibitem{Lamp86}
A.B. Smith, C.D. Jones, and E.F. Roberts,
\newblock ``Article title,''
\newblock {\em Journal}, vol. 62, pp. 291--294, January 1920.

\bibitem{C2}
C.D. Jones, A.B. Smith, and E.F. Roberts,
\newblock ``Article title,''
\newblock in {\em Proceedings Title}. IEEE, 2003, vol.~II, pp. 803--806.

\end{thebibliography}


\begin{thebibliography}{10}
\providecommand{\url}[1]{#1}
\csname url@samestyle\endcsname
\providecommand{\newblock}{\relax}
\providecommand{\bibinfo}[2]{#2}
\providecommand{\BIBentrySTDinterwordspacing}{\spaceskip=0pt\relax}
\providecommand{\BIBentryALTinterwordstretchfactor}{4}
\providecommand{\BIBentryALTinterwordspacing}{\spaceskip=\fontdimen2\font plus
\BIBentryALTinterwordstretchfactor\fontdimen3\font minus
  \fontdimen4\font\relax}
\providecommand{\BIBforeignlanguage}[2]{{%
\expandafter\ifx\csname l@#1\endcsname\relax
\typeout{** WARNING: IEEEtran.bst: No hyphenation pattern has been}%
\typeout{** loaded for the language `#1'. Using the pattern for}%
\typeout{** the default language instead.}%
\else
\language=\csname l@#1\endcsname
\fi
#2}}
\providecommand{\BIBdecl}{\relax}
\BIBdecl

\bibitem{wang2017tacotron}
Y.~Wang, R.~Skerry-Ryan, D.~Stanton, Y.~Wu, R.~J. Weiss, N.~Jaitly, Z.~Yang,
  Y.~Xiao, Z.~Chen, S.~Bengio \emph{et~al.}, ``Tacotron: Towards end-to-end
  speech synthesis,'' \emph{arXiv preprint arXiv:1703.10135}, 2017.

\bibitem{sotelo2017char2wav}
J.~Sotelo, S.~Mehri, K.~Kumar, J.~F. Santos, K.~Kastner, A.~Courville, and
  Y.~Bengio, ``Char2wav: End-to-end speech synthesis,'' 2017.

\bibitem{ping2017deep}
W.~Ping, K.~Peng, A.~Gibiansky, S.~O. Arik, A.~Kannan, S.~Narang, J.~Raiman,
  and J.~Miller, ``Deep voice 3: Scaling text-to-speech with convolutional
  sequence learning,'' \emph{arXiv preprint arXiv:1710.07654}, 2017.

\bibitem{shen2018natural}
J.~Shen, R.~Pang, R.~J. Weiss, M.~Schuster, N.~Jaitly, Z.~Yang, Z.~Chen,
  Y.~Zhang, Y.~Wang, R.~Skerrv-Ryan \emph{et~al.}, ``Natural tts synthesis by
  conditioning wavenet on mel spectrogram predictions,'' in \emph{2018 IEEE
  International Conference on Acoustics, Speech and Signal Processing
  (ICASSP)}.\hskip 1em plus 0.5em minus 0.4em\relax IEEE, 2018, pp. 4779--4783.

\bibitem{zen2009statistical}
H.~Zen, K.~Tokuda, and A.~W. Black, ``Statistical parametric speech
  synthesis,'' \emph{speech communication}, vol.~51, no.~11, pp. 1039--1064,
  2009.

\bibitem{wang2018style}
Y.~Wang, D.~Stanton, Y.~Zhang, R.~Skerry-Ryan, E.~Battenberg, J.~Shor, Y.~Xiao,
  F.~Ren, Y.~Jia, and R.~A. Saurous, ``Style tokens: Unsupervised style
  modeling, control and transfer in end-to-end speech synthesis,'' \emph{arXiv
  preprint arXiv:1803.09017}, 2018.

\bibitem{zhang2019learning}
Y.-J. Zhang, S.~Pan, L.~He, and Z.-H. Ling, ``Learning latent representations
  for style control and transfer in end-to-end speech synthesis,'' in
  \emph{ICASSP 2019-2019 IEEE International Conference on Acoustics, Speech and
  Signal Processing (ICASSP)}.\hskip 1em plus 0.5em minus 0.4em\relax IEEE,
  2019, pp. 6945--6949.

\bibitem{NIPS20188206}
\BIBentryALTinterwordspacing
S.~Arik, J.~Chen, K.~Peng, W.~Ping, and Y.~Zhou, ``Neural voice cloning with a
  few samples,'' in \emph{Advances in Neural Information Processing Systems
  31}, S.~Bengio, H.~Wallach, H.~Larochelle, K.~Grauman, N.~Cesa-Bianchi, and
  R.~Garnett, Eds.\hskip 1em plus 0.5em minus 0.4em\relax Curran Associates,
  Inc., 2018, pp. 10\,019--10\,029. [Online]. Available:
  \url{http://papers.nips.cc/paper/8206-neural-voice-cloning-with-a-few-samples.pdf}
\BIBentrySTDinterwordspacing

\bibitem{valentini2016investigating}
C.~Valentini-Botinhao, X.~Wang, S.~Takaki, and J.~Yamagishi, ``Investigating
  rnn-based speech enhancement methods for noise-robust text-to-speech.'' in
  \emph{SSW}, 2016, pp. 146--152.

\bibitem{kawahara1999restructuring}
H.~Kawahara, I.~Masuda-Katsuse, and A.~De~Cheveigne, ``Restructuring speech
  representations using a pitch-adaptive time--frequency smoothing and an
  instantaneous-frequency-based f0 extraction: Possible role of a repetitive
  structure in sounds,'' \emph{Speech communication}, vol.~27, no. 3-4, pp.
  187--207, 1999.

\bibitem{hsu2019disentangling}
W.-N. Hsu, Y.~Zhang, R.~J. Weiss, Y.-A. Chung, Y.~Wang, Y.~Wu, and J.~Glass,
  ``Disentangling correlated speaker and noise for speech synthesis via data
  augmentation and adversarial factorization,'' in \emph{ICASSP 2019-2019 IEEE
  International Conference on Acoustics, Speech and Signal Processing
  (ICASSP)}.\hskip 1em plus 0.5em minus 0.4em\relax IEEE, 2019, pp. 5901--5905.

\bibitem{narayanan2013ideal}
A.~Narayanan and D.~Wang, ``Ideal ratio mask estimation using deep neural
  networks for robust speech recognition,'' in \emph{2013 IEEE International
  Conference on Acoustics, Speech and Signal Processing}.\hskip 1em plus 0.5em
  minus 0.4em\relax IEEE, 2013, pp. 7092--7096.

\bibitem{wang2014training}
Y.~Wang, A.~Narayanan, and D.~Wang, ``On training targets for supervised speech
  separation,'' \emph{IEEE/ACM transactions on audio, speech, and language
  processing}, vol.~22, no.~12, pp. 1849--1858, 2014.

\bibitem{xie2019utterance}
W.~Xie, A.~Nagrani, J.~S. Chung, and A.~Zisserman, ``Utterance-level
  aggregation for speaker recognition in the wild,'' in \emph{ICASSP 2019-2019
  IEEE International Conference on Acoustics, Speech and Signal Processing
  (ICASSP)}.\hskip 1em plus 0.5em minus 0.4em\relax IEEE, 2019, pp. 5791--5795.

\bibitem{he2016deep}
K.~He, X.~Zhang, S.~Ren, and J.~Sun, ``Deep residual learning for image
  recognition,'' in \emph{Proceedings of the IEEE conference on computer vision
  and pattern recognition}, 2016, pp. 770--778.

\bibitem{zhong2018ghostvlad}
Y.~Zhong, R.~Arandjelovi{\'c}, and A.~Zisserman, ``Ghostvlad for set-based face
  recognition,'' in \emph{Asian Conference on Computer Vision}.\hskip 1em plus
  0.5em minus 0.4em\relax Springer, 2018, pp. 35--50.

\bibitem{wang2018additive}
F.~Wang, J.~Cheng, W.~Liu, and H.~Liu, ``Additive margin softmax for face
  verification,'' \emph{IEEE Signal Processing Letters}, vol.~25, no.~7, pp.
  926--930, 2018.

\bibitem{valin2018hybrid}
J.-M. Valin, ``A hybrid dsp/deep learning approach to real-time full-band
  speech enhancement,'' in \emph{2018 IEEE 20th International Workshop on
  Multimedia Signal Processing (MMSP)}.\hskip 1em plus 0.5em minus 0.4em\relax
  IEEE, 2018, pp. 1--5.

\bibitem{zhang2018deep}
S.~Zhang, M.~Lei, Z.~Yan, and L.~Dai, ``Deep-fsmn for large vocabulary
  continuous speech recognition,'' in \emph{2018 IEEE International Conference
  on Acoustics, Speech and Signal Processing (ICASSP)}.\hskip 1em plus 0.5em
  minus 0.4em\relax IEEE, 2018, pp. 5869--5873.

\bibitem{battenberg2020location}
E.~Battenberg, R.~Skerry-Ryan, S.~Mariooryad, D.~Stanton, D.~Kao, M.~Shannon,
  and T.~Bagby, ``Location-relative attention mechanisms for robust long-form
  speech synthesis,'' in \emph{ICASSP 2020-2020 IEEE International Conference
  on Acoustics, Speech and Signal Processing (ICASSP)}.\hskip 1em plus 0.5em
  minus 0.4em\relax IEEE, 2020, pp. 6194--6198.

\bibitem{45857}
J.~F. Gemmeke, D.~P.~W. Ellis, D.~Freedman, A.~Jansen, W.~Lawrence, R.~C.
  Moore, M.~Plakal, and M.~Ritter, ``Audio set: An ontology and human-labeled
  dataset for audio events,'' in \emph{Proc. IEEE ICASSP 2017}, New Orleans,
  LA, 2017.

\bibitem{reddy2019scalable}
C.~K. Reddy, E.~Beyrami, J.~Pool, R.~Cutler, S.~Srinivasan, and J.~Gehrke, ``A
  scalable noisy speech dataset and online subjective test framework,''
  \emph{Proc. Interspeech 2019}, pp. 1816--1820, 2019.

\bibitem{le2019sdr}
J.~Le~Roux, S.~Wisdom, H.~Erdogan, and J.~R. Hershey, ``Sdr--half-baked or well
  done?'' in \emph{ICASSP 2019-2019 IEEE International Conference on Acoustics,
  Speech and Signal Processing (ICASSP)}.\hskip 1em plus 0.5em minus
  0.4em\relax IEEE, 2019, pp. 626--630.

\end{thebibliography}

\end{document}